\documentclass[journal=jpcafh,manuscript=article]{achemso}
\usepackage{helvet,times,mathptm,epsfig,amsmath}
\usepackage{graphicx}
\usepackage{achemso,natmove}
\usepackage{epsfig}
\newcommand\beq{\begin{equation}}
\newcommand\eeq{\end{equation}}
\newcommand\bea{\begin{eqnarray}}
\newcommand\eea{\end{eqnarray}}

\fontsize{1in} {1.2in}\selectfont

\title{An Exact Solution of the PPP Model for Correlated Electronic
States of Tetracene and Substituted Tetracene}
\author{Y. Anusooya Pati}
\affiliation[]{Solid State and Structural Chemistry Unit, Indian Institute of Science, Bangalore 560 012, India}
\author{S. Ramasesha}
\email{ramasesh@sscu.iisc.ernet.in}
\affiliation[]{Solid State and Structural Chemistry Unit, Indian Institute of Science, Bangalore 560 012, India}
\date{\today}

\begin{document}

\begin{abstract}

{
Tetracene is an important conjugated molecule for device applications. We
have used the diagrammatic valence bond method to obtain the desired states, 
in a Hilbert space of about 450 million singlets and 902 million triplets. 
We have also studied the donor/acceptor (D/A) substituted tetracenes with
D and A groups placed symmetrically about the long axis of the molecule. In 
these cases, by exploiting a new symmetry, which is a combination of $C_2$ 
symmetry and electron-hole symmetry, we are able to obtain their low-lying 
states. In the case of substituted tetracene, we find that optically allowed 
one-photon excitation gaps reduce with increasing D/A strength, 
while the lowest singlet-triplet gap is only weakly affected. In all 
the systems we have studied, the excited singlet state, $S_1$ is at more 
than twice the energy of the lowest triplet state and the second triplet 
is very close to $S_1$ state. Thus donor-acceptor substituted tetracene could 
be a good candidate in photo-voltaic device application as it satisfies 
energy criteria for singlet fission. We have also obtained the model exact
second harmonic generation (SHG) coefficients using correction vector 
method and we find that the SHG responses increase with the increase 
in D/A strength. 
}
\end{abstract}
\maketitle

\section{Introduction}

There is an increased interest in the study of polycyclic hydrocarbons, 
particularly, tetracene and pentacene since the last decade due to their 
large hole mobility and improved field effect transistor (FET) 
efficiencies \cite{ieee}. They are 
used in Organic Light Emitting diode (OLED) 
applications, in field effect transistors and photovoltaic devices \cite{}.
Metal doped pentacenes and picenes show superconductivity at relatively
high $T_c$ values (above 7K) \cite{picene-nature-sc}. These systems are 
building blocks of graphene and are semiconducting in nature. Organic 
counterpart of inorganic semiconductors are more easy to process and to 
tailor for required applications with easy substitution. Substitution by 
electron donating and  withdrawing groups leads to ambipolar materials 
which are used in organic 
photovoltaic cells \cite{chem-asia}. Longer acenes are found to be less 
stable and hence there are efforts to derivatize the parent tetracene  and
pentacene compounds to make them more soluble and 
stable \cite{yutaka-org-elec}. 
Yutaka et al have synthesized benzopyrazine-fused tetracene compounds and 
found that these compounds  are more photostable and have long wavelength 
absorption.
The major aim is to tune the HOMO - LUMO gap to assist the easy 
flow of positively charged holes and negatively charged electrons, either for
recombination or for charge separation, depending upon the application 
i.e. LEDs or photovoltaics. One way to reduce 
the HOMO-LUMO gap is to increase the conjugation length of the molecule and 
another is to substitute the systems with electron withdrawing and donating 
groups \cite{tetra-chem-rev}.

    A recent paradigm in the field of organic photovoltaics is the fission 
of the photoexcited singlet  into two triplets 
\cite{sf-pope-book,michl-chem-rev,michl-ann-rev}. These triplets generated by fission 
can then undergo dissociation to yield twice the number of charge carriers
that is  produced by singlet dissociation. There are several 
conditions under which this can happen with larger probability. They are 
(i) the energy $E(S_1)$ of the lowest excited singlet state, $S_1$, is 
greater than or equal to twice the triplet energy ($E(S_1) \ge 2 E(T_1)$, 
(ii) the second triplet state, $T_2$, is above the singlet excited 
state, $S_1$, i.e. ($E(T_2) > E(S_1)$), as this will avoid leaking of the 
$S_1$ state to $T_2$ via intersystem crossings and (iii) $E(T_1)$ should be 
at least 1 eV as otherwise the operating voltage of the OPVC will drop, 
resulting in lower efficiency. Polycrystalline tetracene and pentacene 
molecules have been explored in this 
context \cite{zimmer-natchem, zimmer-jacs, tetra-sf-natchem, acr-friend}.  
Effect of magnetic field on SF has been studied 
by Bardeen et al \cite{feature-cpl}.

 There are several theoretical studies of these systems using both 
semi-empirical and {\it ab-initio} methods and also by the density matrix 
renormalization group (DMRG) method \cite{joc-wiberg, oligo-acene-jcp, 
acene-tddft-jcp,acene-prb1, acene-prb2}. The energetics and structural 
parameters of acene series and their analogues - phenanthrene series, has 
been studied by Wiberg using the DFT method \cite{joc-wiberg}. They analyze 
their results by studying quantities like resonance energy, ionization potential 
and 
$\sigma-$ and $\pi-$ bond indices. Heinze et al have studied the excitation 
energies and oscillator strengths of acenes using their method based on 
time dependent density functional theory (TDDFT) \cite{acene-tddft-jcp}. 
Excitation energies of longer acenes are studied by Kadantsev et al within 
TDDFT method, both in the singlet and triplet manifold \cite{oligo-acene-jcp}. 
The triplet-triplet transitions were experimentally measured by 
Pavlopoulos \cite{trip-trip-spectra}. Kaur et al studied the effect of 
substituent on the HOMO-LUMO gaps on pentacene \cite{kaur-jacs}. Aldehyde 
substituted oligoacenes were studied for their enhanced first order 
hyperpolarizabilities using hyper Rayleigh scattering technique \cite{hrs-cpl}. 
The effect of donor-acceptor groups on the first order polarizabilities 
of substituted oligoacenes were  studied within 
AM1/TDHF method \cite{costa-jmatsci}.

       The DFT method is basically a ground state method and is helpful for 
obtaining ground state properties such as molecular geometries. Although the 
TDDFT method, in principle,  can provide excited state information, it 
suffers from the severe drawback of lack of reliable functionals. Indeed 
both these methods are similar in line with the well established Hartree-Fock 
(HF) and TDHF methods which include mean exchange and correlation potentials.
All the above methods include both coulomb and exchange 
correlation, but only at the mean-field level. For obtaining electronic
excited state properties for conjugated systems, it has been demonstrated 
that using a model of $\pi$- electrons and treating electron-electron 
interaction with a very high level theory gives accurate excited states 
and their properties \cite{sr-rev}. In this spirit, we have employed the Pariser-Parr-Pople (PPP) model for describing 
 $\pi$ electrons. The PPP model includes long-range electron-electron 
interactions and is suited for semiconducting systems.

           In this paper, we have studied tetracene and its donor-acceptor
substituted compounds by solving the PPP model exactly using a diagrammatic 
valence bond (DVB) approach \cite{sr-ijqc,soos-vb}. Tetracene molecule consists of 
18 $\pi-$ electrons delocalized over the 18 Carbon atoms of tetracene. The 
full configuration  space of tetracene spans over 0.9 billion configurations 
for triplets, without taking into account the three fold spin degeneracy of 
triplets and extends DVB calculations to nearly a billion valence bond 
functions. We have computed excitation energies 
of these compounds and analyzed their oscillator strengths and geometries 
both in the ground state and excited states. Besides, we have also 
explored the triplet states of these systems in the context of singlet 
fission. We have  obtained the model exact SHG response of these systems
using the correction vector techniques. In what 
follows, we give a brief introduction to the DVB method and model 
Hamiltonian used, followed by results and discussion. 

\section{Methodology}
The PPP model assumes  $\sigma-\pi$ separability and considers a   single 
$p_{z}$ orbital at each carbon site, for tetracene this translates to a 
problem of 18 electrons on 18 site. The PPP Hamiltonian in second 
quantization notation, with $a^\dagger_{i\sigma}$ ($a_{i\sigma}$) creating 
(annihilating) an electron with spin $\sigma$ in orbital (site) $i$ with 
$n_i$ being corresponding occupation number operator, is given by 

\begin{eqnarray}
 H & = & \sum_{<i,j>\sigma}t_{ij} ( a^\dagger_{i\sigma}a_{j\sigma} +
 a^\dagger_{j\sigma}a_{i\sigma}) +  \\ \nonumber
& & + \sum_{i} \epsilon_i n_i + \frac{1}{2} \sum_{i} U_i n_i (n_i-1) 
\\ \nonumber
& & + \sum_{i > j} V_{ij} (n_i-z_i) (n_{j}-z_j)  
\end{eqnarray}

\noindent

\begin{figure}
\begin{center}
\includegraphics[height=6.0cm,width=15cm]{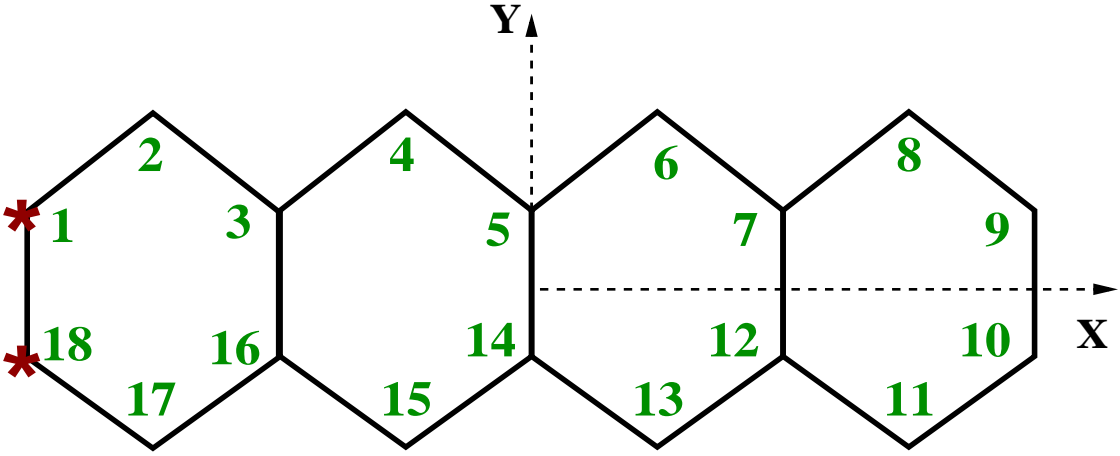} \\
\caption{Schematic structure of tetracene. The sites $1$ and $18$ are
substituted by donor and acceptor ($+$ $\epsilon$ and $-$ $\epsilon$),
respectively. }
\label{mol-struct-fig}
\end{center}
\end{figure}

The first term in the Hamiltonian corresponds to the kinetic energy. 
$t_{ij}$s are the resonance/hopping (transfer) integrals between bonded 
carbon sites $i$ and $j$. The second term corresponds to the site energy 
with $\epsilon_i$ being the orbital energy of the $p_z$ orbital on 
the $i^{th}$ carbon atom. $U_i$s are the on-site electron-electron repulsion 
parameter (the Hubbard parameter) at site $i$ and $V_{ij}$s are  intersite 
electron-electron repulsion parameters between sites $i$ and $j$.  $z_i$ is the local chemical potential at site $i$ 
which is $1$ for carbon $\pi-$ orbitals. The parameters
$t_{ij}$s are taken as $-2.4$ eV and $U_i$s are $11.26$ eV and $V_{ij}$s are 
obtained using the Ohno~\cite{ohno} interpolation formula, 

{\bf 
\begin{equation}
V_{ij}=\frac{U_i} {\sqrt{(1.0 + 0.6117 r_{ij}^2})}  
\label{ohno-eqn}
\end{equation}}
 
\noindent
where $r_{ij}$ is the intersite distance  in \AA.
Site energy $\epsilon$ is taken as zero for unsubstituted $\rm C$ atoms. We 
have mimicked the effect of substitution by donors or acceptors at a site $i$
 by changing the site energies of the carbon atoms at these sites. Donor site
has a +ve site energy while an acceptor site has a negative site energy. We 
have assumed equal strength of donors and acceptors and varied the magnitude 
of site energy $ |\epsilon|$ from $ 2.0$ to $4$ eV. We have introduced the 
substituents such that they are at sites related by the $C_2$ axis along the 
length of the molecule as shown in \ref{mol-struct-fig}.

The unsubstituted tetracene molecule, has spatial symmetry ($C_2$) and  
electron-hole symmetry ($e$-$h$) assuming all carbon sites are identical,
leading to an Abelian group  of 4 elements. Both these symmetries are broken, 
when we introduce donor and acceptors in the system. However, if the donor 
and acceptors are at sites related by $180 ^{\circ}$  rotation about 
the long axis of 
the molecule and if the magnitude of donor-acceptor strengths  are the same, we 
will still retain the symmetry corresponding to $C_2 \times e$-$h$. This can be 
seen by noting that the $C_2$ symmetry  interchanges sites $j$ and $(N+1)-j$, 
where $N=18$ in tetracene. The $e$-$h$ symmetry transforms the creation 
operator $a^{\dagger}_i$  at site $i$ to annihilation operator $a_i$, while 
at site $(N+1)-i$  it interchanges  $a^{\dagger}_{(N+1)-i} = -a_{(N+1)-i}$ since
sites $i$ and $(N+1)-i$  belong to different sublattices. At half-filling the
interaction terms and transfer terms in the substituted tetracene are the 
same as those in the unsubstituted tetracene and hence their invariance 
under  $C_2 \times e$-$h$ operator is well established. The only additional term
is the site energy term $\sum_i \epsilon_i n_i$ and for substitutions at sites
$j$ and $(N+1)-j$, the summation can be written explicitly as 
$\epsilon_j n_j + \epsilon_{(N+1)-j} n_{(N+1)-j}$. Since we impose equal donor and
acceptor strengths $\epsilon_j = - \epsilon_{(N+1)-j} $
and site energy terms reduce to 
$\epsilon_j n_j - \epsilon_{(N+1)-j} n_{(N+1)-j} $.
Operating on this by $e$-$h$ leads to $-\epsilon_j n_j + 
\epsilon_{(N+1)-j} n_{(N+1)-j} $ and
$C_2$ operation on this term restores the original  term in the Hamiltonian.
By employing this symmetry for symmetrically substituted donor-acceptor groups
in tetracene, we can reduce the Hilbert space dimension, approximately, 
by half. The largest subspace we have dealt with corresponds to the triplet 
space of tetracene with symmetric substitution which has a dimension
of $\approx$ 0.45 billion. The valence bond (VB) technique for solving the
PPP Hamiltonian is followed along the lines described in earlier work\cite{sr-ijqc,soos-vb}.

\noindent

\section{Results and Discussion}
           
\subsection{Singlet State Properties}
In the case of tetracene, we have obtained a few low-lying singlet
and triplet states in the $A^+$ and $B^-$ subspaces. In the case  of 
substituted tetracene, we have computed a few low-lying states in 
the $\Sigma$ and $\tau$ subspaces where $\Sigma$ corresponds to even 
subspace and $\tau$ to the odd subspace, under $C_2 \times e$-$h$. 
In substituted tetracenes, it is worth noting that the optical transitions 
are allowed between states within the same 
subspaces i.e. $\Sigma \rightarrow \Sigma$ or $\tau \rightarrow \tau$, 
besides the usual $\Sigma \rightarrow \tau $ transitions. The 
$\Sigma \rightarrow \Sigma$ transitions are polarized along the 
short-axis (Y-axis) of the molecule while the $\Sigma \rightarrow \tau$ 
transitions are polarized along the long axis (X-axis) of the molecule. 

   Tetracene molecule has $D_{2h}$ symmetry. We have assumed
planar geometry and have ignored the Hydrogen atoms. Therefore, the symmetry
reduces to $D_2$, since $C_2(Z)$ is the same as inversion, for a planar
molecule. The states of tetracene can therefore be classified as $A^+$,
$B_1^+$, $B_2^+$, $B_3^+$,  $A^-$,  $B_1^-$, $B_2^-$ and $B_3^-$ where
the superscripts $+$ and $-$ refer to the electron-hole symmetry,
$+$ for even space and $-$ for odd space, representing covalent and ionic
spaces. Since we have not used the $C_2$ symmetry along the Y-axis, to 
uniquely assign the state lables,  we have used the direction
of polarization of the transition dipole between the ground state and 
excited states. The transition to $B_1$ is Z-polarized and will be disallowed 
as the molecule is in the XY-plane. Transitions to $B_2$ states are $Y-$ 
polarized and to $B_3$ states are $X-$ polarized. 

\begin{center}
\begin{table} 
\setlength{\tabcolsep}{2.0pt}
\begin{tabular}{|c|c|l|l|l|l|l|l|l|} \hline
$\epsilon$ & &  \multicolumn{7}{|c|} {Excited State index} \\ \cline{3-9} 
    &          & \hspace{0.8cm} 1   &  \hspace{0.7cm} 2 ~~& \hspace{0.8cm} 3 ~ & 
\hspace{0.7cm}  4 ~~ & \hspace{0.8cm}  5 ~ & \hspace{0.8cm}  6& \hspace{0.8cm} 7  \\ \hline
    &          &       &       &       &      &      &       &      \\
0.0 &   gap    & ~ 3.18 ($B^-_2$) ~ & ~~ ---~~  & ~ 3.59 ($B^+_1$)~ &
 ~~ 3.97 $^\star$ ($B_3$) ~~  &~ 4.14 ($B^-_1$)~ & ~~ 4.95 ($B^-_3$) ~ &
 ~ 4.99 ($B^-_3$) ~   \\
    & $\mu_x $ & ~ 0.00 ~ & ~~ --- ~~ & ~ 0.00~ & ~~ 0.04  ~~ &
    ~ 0.00~ & ~ 11.45~ & ~ 1.32~  \\
    & $\mu_y $ & ~ 3.74 ~ & ~~ --- ~~ & ~ 0.00~ & ~~ --- ~~  &~ 0.00~ 
    & ~~~ 0.00~ & ~ 0.00~  \\
    &  & ~~[2.71]\cite{bier-schm-jacs}~~ & ~~ ---~~  & ~~ ---~~  
    & ~~  [3.32]\cite{bier-schm-jacs}~~ & ~~ ---~~ 
    & ~~ [4.52] \cite{bier-schm-jacs}~~ & ~~ ---~~   \\ 
    &  & ~~[2.63]\cite{tetra-dimer}~~ &~~ ---~~   &~~ ---~~   
    & ~~ ---~~       & ~~ ---~~ &~~ [4.51] \cite{tetra-dimer}~~ &   \\
    &          &       &       &       &      &      &       &       \\

2.0 &   gap    & ~ 2.72~($\Sigma$) & ~ 3.06~($\Sigma$)~ &~ 3.51~($\tau$) 
    &~ 4.31~($\tau$) & ~ 4.71~($\tau$)~ & ~~ 4.91~($\tau$) &~  5.46~($\tau$)  \\
    & $\mu_x $ & ~ 0.00~ & ~ 0.00~ &~ 0.69 ~ & ~ 1.34~ 
    & ~ 1.86~ & ~ 11.05~ &~  1.48~  \\
    & $\mu_y $ & ~ 3.27~ & ~ 0.13~ &~ 0.00 ~ & ~ 0.00~ 
    & ~ 0.00~ & ~~~ 0.00~ &~  0.00~  \\
    &         &       &       &       &       &      &       &      \\

3.0 &   gap    & ~ 2.45~($\Sigma$) & ~ 3.06~($\Sigma$)~ &~ 3.41~($\tau$) ~
    & ~ 4.24~($\tau$)~ & ~ 4.65~($\tau$) ~& ~~~ 4.86~($\tau$)~ &~ 5.35~($\tau$)~   \\
    & $\mu_x $ & ~ 0.00~ & ~ 0.00~ &~ 0.89 ~ & ~ 1.63~ 
    & ~ 1.84~ & ~ 10.64~ &~ 2.82~  \\
    & $\mu_y $ & ~ 3.56~ & ~ 0.58~ &~ 0.00 ~ & ~ 0.00~ 
    & ~ 0.00~ & ~~~ 0.00~ &~ 0.00~  \\
    &          &       &       &       &       &       &       &      \\ 

4.0 &   gap    & ~ 2.20~($\Sigma$) & ~ 3.04~($\Sigma$)~ &~ 3.30 ~($\tau$) 
    & ~ 4.15~($\tau$) & ~ 4.59~($\tau$) & ~~~ 4.79 ($\tau$)~
    & ~ 5.22~($\tau$) ~   \\ 
    & $\mu_x $ & ~ 0.00~ & ~ 0.00~ &~ 0.94 ~ & ~ 1.59~ 
    &  ~ 1.31~ & ~ 10.05~ &~ 4.24~  \\
    & $\mu_y $ & ~ 3.75~ & ~ 0.90~ &~ 0.00 ~ & ~ 0.00~ 
    & ~ 0.00~ & ~~~ 0.00~ &~ 0.00~  \\
\hline
\end{tabular}
\caption{Low-lying singlet-singlet excitations in tetracene as a function of 
site energy, $\epsilon$. Energies are in eV and transition dipole moments are 
in Debye. $\Sigma$ corresponds to the even space and $\tau$ to odd space under
$C_2 \times e$-$h$ symmetry. The number with $\star$ is obtained by 
introducing a small site energy at inequivalent sites of 
tetracene \cite{sr-soos-napth}.}
\label{tab-sing}
\end{table}
\end{center}

   In \ref{tab-sing} we present the low-energy excitations of tetracene 
($\epsilon=0$) and substituted tetracene ($\epsilon \ne 0$). 
The  lowest singlet excitation is at 3.04 eV to $A^+$ state which is a 
two-photon state. 
In the polyacene series, it is known that two-photon state is above
one photon state for tetracene, while for pentacene, the two photon state
is below the one photon state \cite{acene-prb2}. Our results seem to
 indicate that even for tetracene and for pentacene, the two photon state 
is below the one photon state. The energy gap of 0.14 eV between the two 
states is too small to definitely state that the two photon state is below 
one photon state in the crystal as intermolecular interactions will red shift 
the one photon state more than the two photon state, the former being
more ionic. 

Optically allowed excitations are to the $B^-_2$  state at 3.18 eV (weaker) 
and the $B^-_3$ state at 4.95 eV (stronger). Both these excitations are 
blue-shifted with respect  to the experimental values by $\approx 0.5$ eV 
\cite{bier-schm-jacs,tetra-dimer}. The excitation around 
$3.3$ eV is very very weak and to observe this peak we need to take into 
account the inequality of $C$ sites in tetracene \cite{sr-soos-napth}. 
If we take  a slightly negative site energy ($\epsilon = -0.15~~ eV$) 
for $C_3$, $C_5$, $C_7$, $C_{12}$, $C_{14}$ and $C_{16}$ and calculate
the energy spectrum, we observe a weak peak at $3.97~ eV$ (with a transition
dipole of 0.04 Debye),
which is $0.65 ~eV$ higher than the experimental value.

On introducing substitution, the strong optically allowed state red shifts
progressively from 3.18 eV to 2.72 eV,  2.45 eV and  2.20 eV for 
$\epsilon = 2.0$,  $3.0$  and $4.0$ eV, respectively. All these excitations 
are short axis polarized. The next strongest optical excitation is at a gap 
of $4.95$, $4.91$, $4.86$ and $4.79$ eV for $\epsilon = 0.0$, $2.0$, 
$3.0$ and $4.0$ eV, respectively.  All these excitations are long axis 
polarized and show a smaller red shift with increasing strength of 
substitution. For unsubstituted tetracene, the third optically allowed 
excitation is nearly degenerate with the second excitation with a small 
transition dipole along the long axis of tetracene. Upon substitution, this 
state gets blue-shifted. The excitation energy of this level reduces with the
increasing $\epsilon$ value, while the transition dipole moment 
increases, retaining its direction of polarization.

\subsection{Charge Density}
\begin{figure}
\begin{center}
\includegraphics[height=6.0cm,width=12cm]{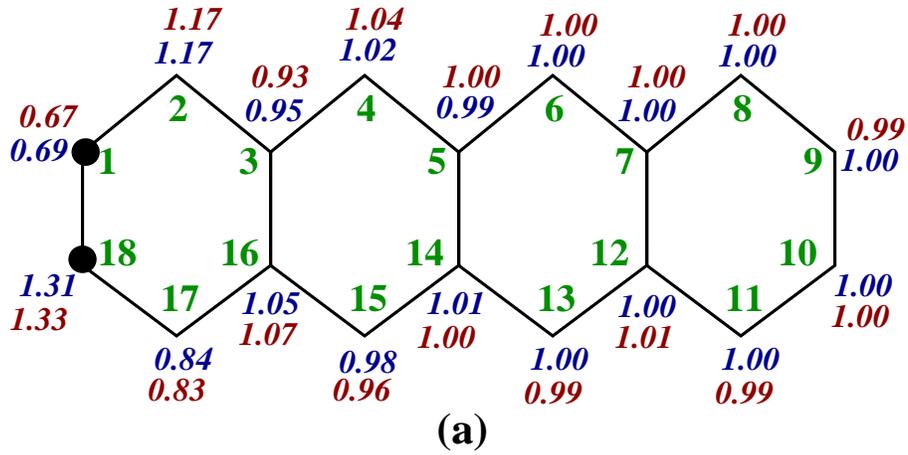} \\
\vspace{0.5cm}
\includegraphics[height=6.0cm,width=12cm]{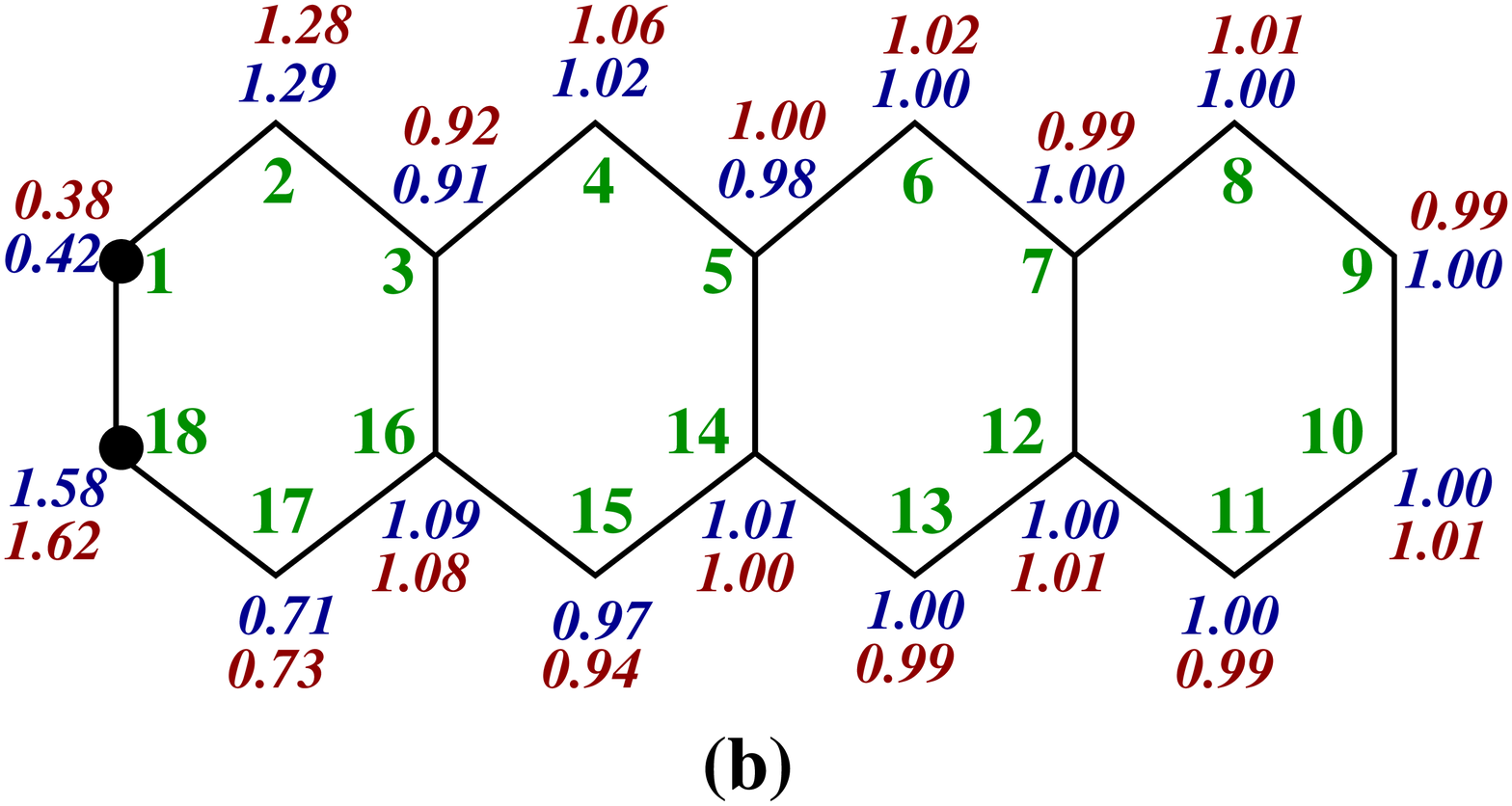} \\
\vspace{0.5cm}
\end{center}
\caption{Charge densities for ground state (in blue) and dipole 
allowed vertically excited state ($\tau_4$, in red) as a function of 
site energy, $\epsilon$. (a) for $\epsilon = 2.0$ eV and (b) 
for $\epsilon = 4.0$ eV. Numbers inside the ring in green represent 
the site/orbital indices.}
\label{chr-fig}
\end{figure}

We have computed the charge density and bond orders for these systems both
in the ground state and excited states, which have significant transition
dipole moments to the ground state. Because of the $e$-$h$ symmetry, 
the charge density at every carbon site is 1 for an unsubstituted molecule 
both in the ground state and excited states. In \ref{chr-fig}, we have given
the charge densities for two different site energies, $\epsilon = 2.0$ eV and
$\epsilon = 4.0$ eV. For site energy, $\epsilon = 3.0$ eV, we have given 
the charge density data in 
supporting information. At the site of the substitution, the charge density 
difference is large and
it slowly varies alternately along the long - axis of the molecule and reaches
the value of 1 away from the sites of substitution. This limiting value is
attained over shorter distances from the substituted sites for weaker 
donor-acceptor strengths. For example, the effect of substitution is seen 
till the second ring in the case of $\epsilon = 2.0$, whereas, it is spread 
upto the third ring for $\epsilon = 4.0$. The magnitude of difference in 
charge density varies almost linearly with the strength of the site energy. 
Another interesting observation is that the sum of the charge densities at 
sites related by $C_2$ symmetry about the long-axis is 2.0, which is a 
consequence of the $C_2 \times e$-$h$ symmetry.

\begin{figure}
\begin{center}
\vspace{0.5cm}
\includegraphics[height=6.0cm,width=12cm]{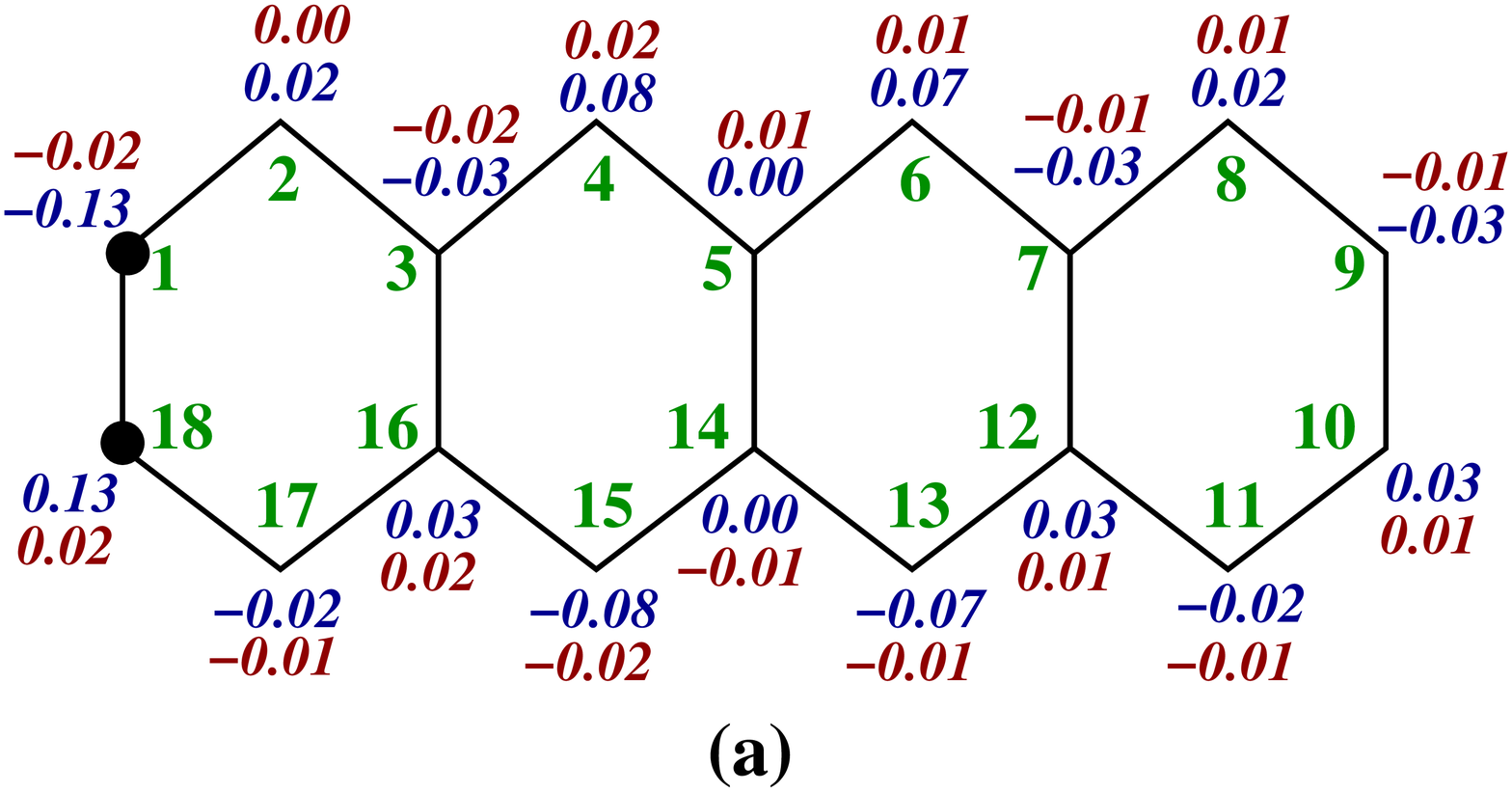} \\
\vspace{0.5cm}
\includegraphics[height=6.0cm,width=12cm]{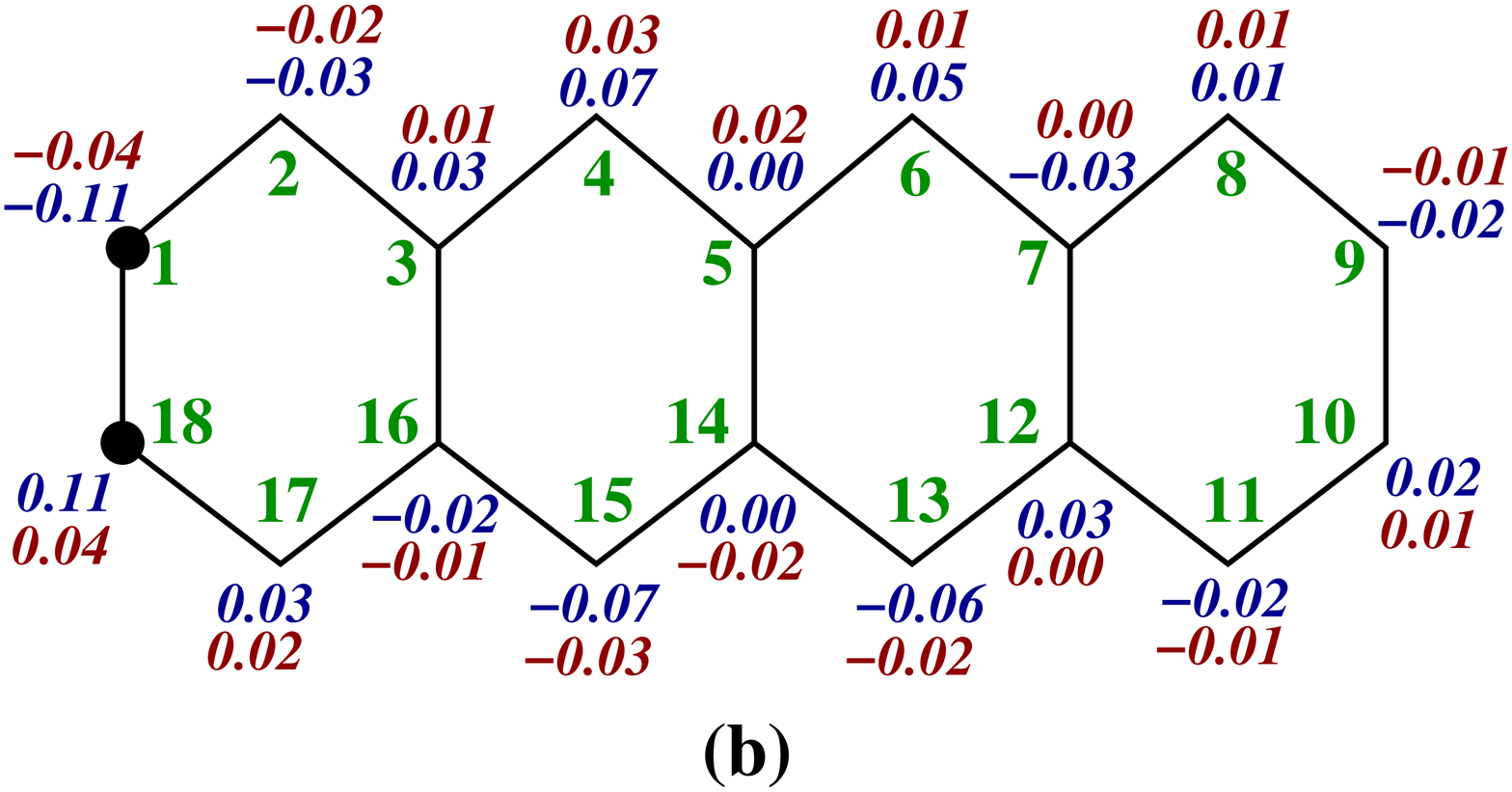} \\
\vspace{0.5cm}
\end{center}
\caption{Difference in charge densities from ground state to vertically 
excited states in even space ($\Sigma_2$, in blue) and odd space 
($\tau_4$, in red), as a function of site energy, $\epsilon$. (a) for 
$\epsilon = 2.0$ eV and (b) for $\epsilon = 4.0$ eV. Numbers inside the 
ring in green represent the site/orbital indices.}
\label{chrdif-fig}
\end{figure}

     In \ref{chrdif-fig}, we have given the difference in charge density 
between the ground state and excited states which are dipole allowed. As 
expected, the difference is more at the substituted sites and the magnitude 
of difference is same at the sites related by $C_2$ symmetry, which is again
a consequence of  the $C_2 \times e$-$h$ symmetry. In the excited 
states, the non-zero difference extends to the farthest sites from the 
substitution sites. This is in contrast to the ground state charge 
distribution, which is more localized closer to the site of substitution. 
The magnitude of the difference is larger for the state which has non-zero 
transition dipole moment along the Y- axis.

\subsection{Bond Orders in Singlet states}

Bond order $p_{ij}$ of a bond between sites $i$ and $j$ in the state $|\psi>$ is defined as 
\begin{equation}
p_{ij} = -\frac{1}{2} <\psi | \sum_{\sigma} {a^{\dagger}_{i \sigma} a_{j \sigma} + h.c. |\psi>}
\end{equation}

\begin{figure}
\begin{center}
\vspace{0.5cm}
\includegraphics[height=6.0cm,width=12cm]{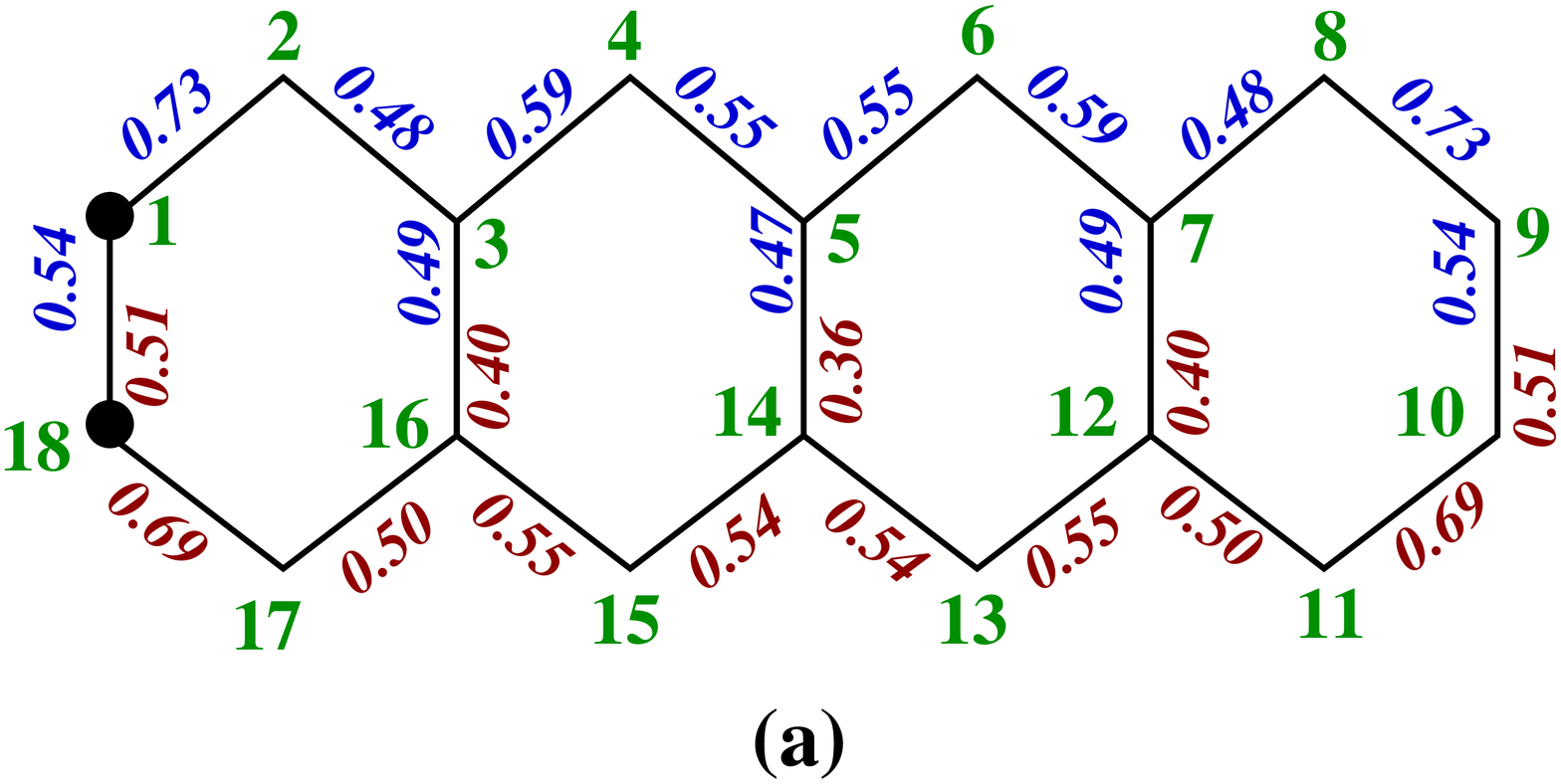} \\
\vspace{0.5cm}
\includegraphics[height=6.0cm,width=12cm]{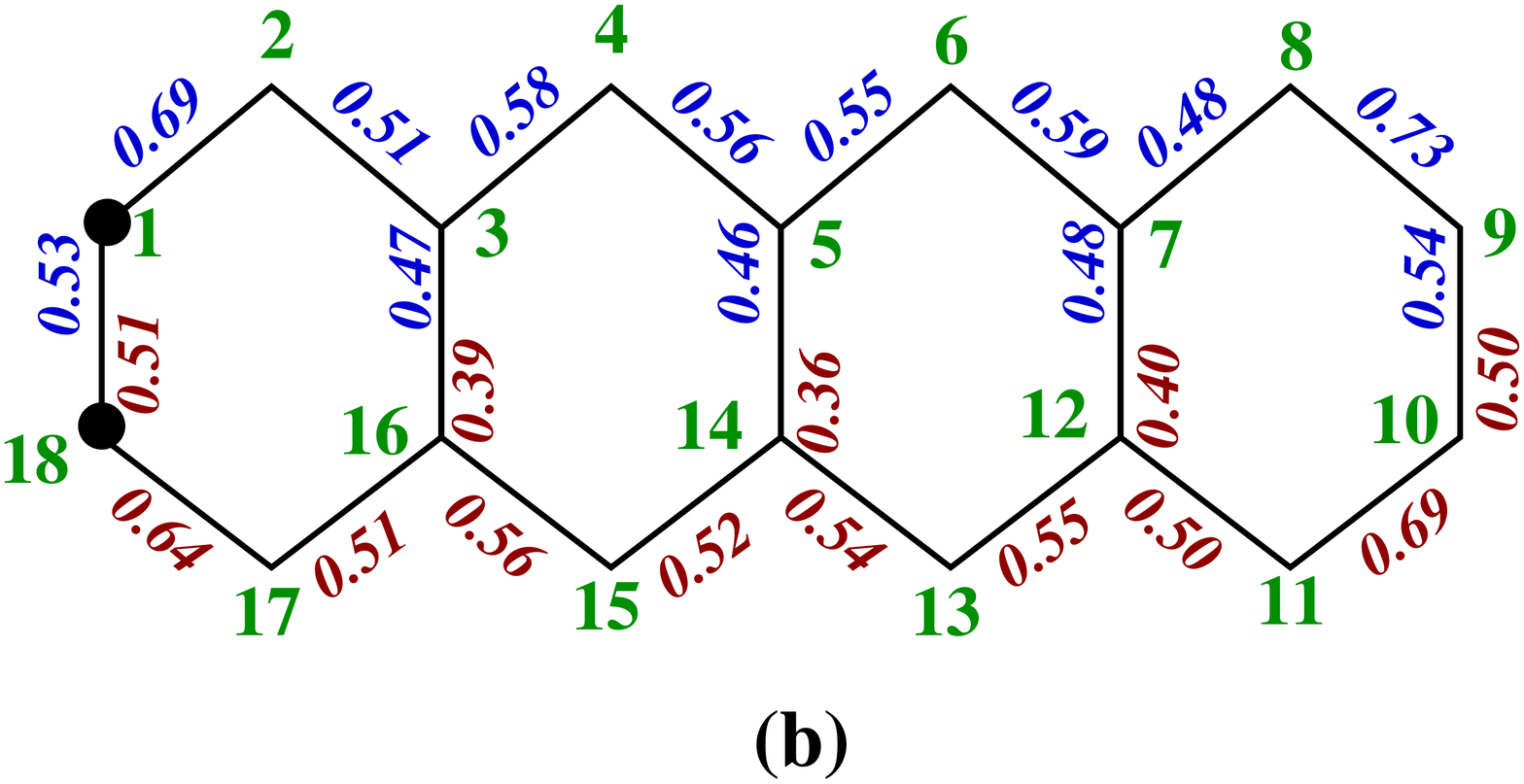} \\
\vspace{0.5cm}
\includegraphics[height=6.0cm,width=12cm]{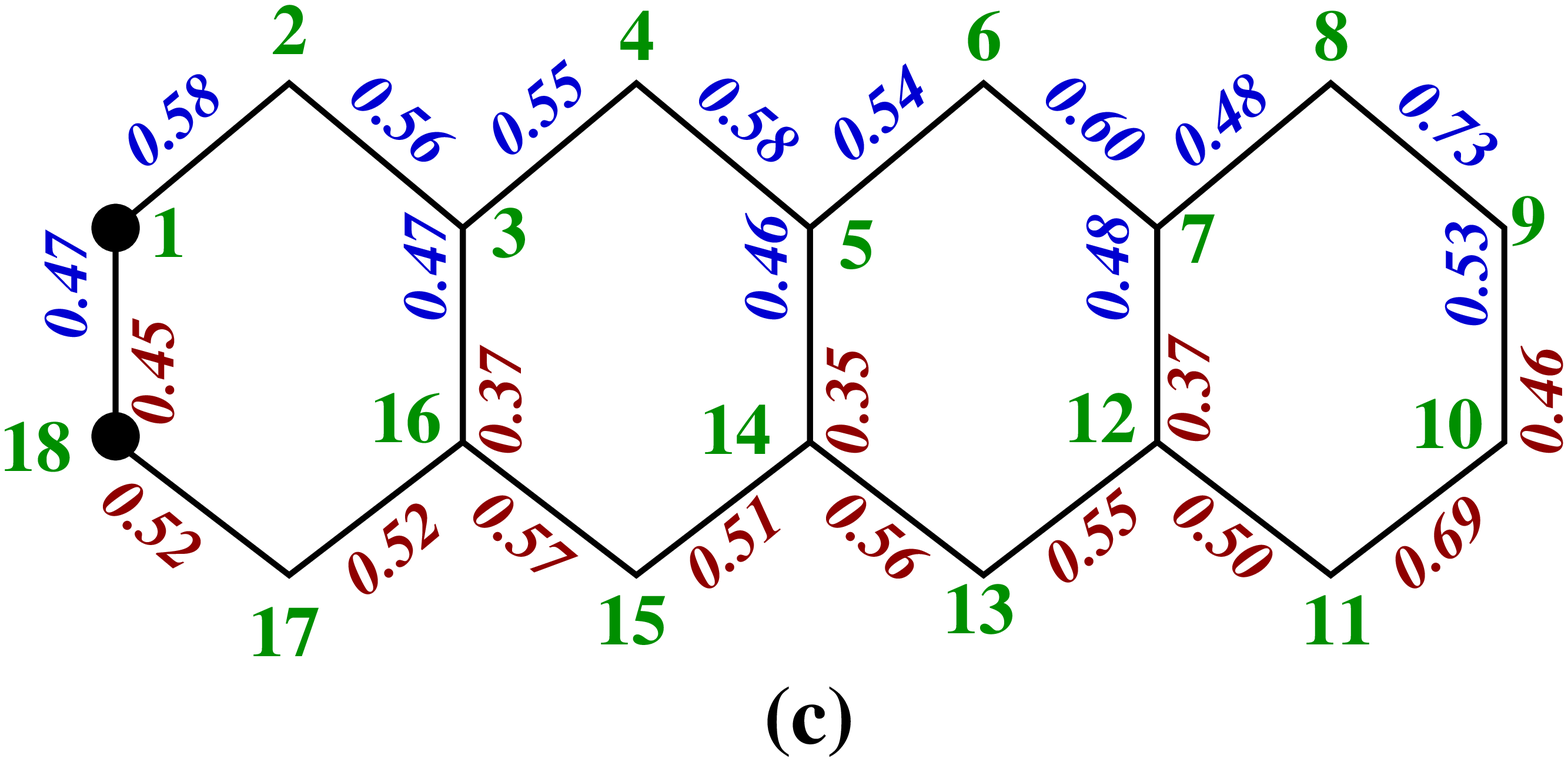} \\
\vspace{0.5cm}
\end{center}
\caption{Bond Order for  ground state ($\Sigma_1$, in blue) and optically 
allowed state ($\tau_4$, in red), as a function of site energy, $\epsilon$. 
(a) for $\epsilon = 0.0$ eV, (b) for $\epsilon = 2.0$ eV and (c) for 
$\epsilon = 4.0$ eV.  Numbers inside the ring in green represent the 
site/orbital indices.}
\label{bnd-fig}
\end{figure}

A larger bond order implies that at equilibrium, the bond would contract 
while  smaller bond order implies the tendency for the bond to elongate.
 At equilibrium, all bond orders will be proportional to their respective 
bond lengths, with the same proportionality constant. Thus, a study of 
the bond order in different states gives an idea of the equilibrium geometry. 
In \ref{bnd-fig}, we present the bond orders for the ground state (numbers in 
blue) and excited states ($\tau_4$, numbers in red) for tetracene and 
substituted tetracenes. In the ground state, outer bonds show strong 
bond alternation while the inner bonds ($p_{4,5}$ and $p_{5,6}$) 
tend to be uniform.  The rung bonds are 
weaker and of similar magnitude except in outer most rings (bonds $p_{1,18}$ 
and $p_{9,10}$). Our bond order patterns compare well with the bond order 
patterns obtained by Wiberg who computed the Fulton $\pi-$ bond indices for 
tetracene\cite{joc-wiberg}. The effect of substitution on bond order is more 
pronounced near the site of substitution, similar to the behavior of charge 
density. 

Upon excitation, the stronger bonds become weaker and vice versa, along the 
chain and the rung bonds become even more weaker. Moreover, the rung bonds show 
larger variation compared to the bonds along the chain. In the case of 
substituted tetracenes, bond order variation is also more localized when 
the site energy is small and is delocalized when it is large. In all the 
cases, the excited state geometry is more enlarged than the ground state, 
since the magnitude of elongation of double bond is larger than the 
contraction of the single bond. The effect of substitution on bond order 
is more pronounced near the site of substitution, 
similar to that observed for charge density behavior. 

\section{Properties of Triplet States}

We have computed several excited triplet states in different symmetry subspaces 
of tetracene and substituted tetracene. Triplet state energies and 
triplet-triplet transition dipoles are presented in \ref{tab-trip}. In 
unsubstituted tetracene, the lowest T-T transition that is optically allowed 
from $T_1$  is at 3.07 eV and the transition is long-axis polarized. 
All the triplet states below this state have no transition dipole for 
optical excitation.
Experimentally two nearly degenerate peaks are observed at 
2.61 and 2.70 eV\cite{trip-trip-spectra}. We observe two more weaker peaks 
at 3.82 (Y-axis polarized) and 4.50 (X-axis polarized). These peaks are 
comparable to the experimental peaks at 3.97 eV and 4.36 eV with different 
polarization axes found experimentally \cite{trip-trip-spectra}.

On introducing substitution, the lowest T-T transition from $T_1$ is 
at $\approx $ 1.05 eV, independent of the strength of substitution, but
the transition dipole increases with increase in substitution strength. This 
state is not dipole connected to $T_1$ in unsubstituted tetracene. 
We observe many weaker peaks ($T_2$ to $T_7$) below 3.07 eV of unsubstituted 
tetracene, all of which are dipole connected to $T_1$ state. The excitations to 
$T_2$, $T_3$ and $T_5$ are short-axis polarized for $\epsilon = 2.0$ and
$T_2$, $T_3$ and $T_6$ are short-axis polarized for $\epsilon = 3.0$ and
$\epsilon = 4.0$ eV.  The remaining transitions are long-axis polarized. 
The transition to $T_2$, $T_3$ and $T_7$ show increase in transition dipole 
with increasing $\epsilon$. There seems to be level crossings with $\epsilon$, 
for states $T_4$, $T_5$ and $T_6$. For example, $T_5$ and $T_6$ seem to cross 
for $\epsilon > 2.0 $ eV. The strongly allowed T-T transition, $T_8$, in 
unsubstituted tetracene becomes progressively weakly allowed, as the 
substitution strength is increased. 

\begin{center}
\begin{table} 
\setlength{\tabcolsep}{2.0pt}
\begin{tabular}{|c|c|l|l|l|l|l|l|l|} \hline
$\epsilon$ & &  \multicolumn{7}{|c|} {Energies of excited states } \\ \cline{3-9} 
    &          & \hspace{0.7cm} 2      &   \hspace{0.7cm} 3    &   \hspace{0.7cm} 4    &
\hspace{0.7cm} 5      & \hspace{0.7cm} 6     &\hspace{0.7cm} 7     & \hspace{0.7cm} 8 \\ \hline
    &          &       &       &       &      &      &     &   \\
0.0 &   gap    & ~ 1.05~ ($B_2$)~ & ~ 1.92~ ($B_2$)~ & ~ 1.95~ ($B_1$) ~
    & ~ 2.46~ ($B_1$) ~& ~ 2.69~ ($B_1$)~ & ~ 2.89~ ($B_1$)~ & ~ 3.07~ ($B_3$)~ \\
    & $\mu_x $ & ~ 0.00 ~ & ~ 0.00~ & ~ 0.00~ & ~ 0.00 ~ & ~ 0.00 ~ &  ~ 0.00~ & ~ 7.02~ \\
    & $\mu_y $ & ~ 0.00 ~ & ~ 0.00~ & ~ 0.00~ & ~ 0.00~ & ~ 0.00 ~ &  ~ 0.00~ & ~ 0.00~ \\
    &          &          &         &         &          &   &         &       \\

2.0 &   gap    & ~ 1.06~ ($\Sigma$) ~& ~ 1.94~ ($\Sigma$) ~ & ~ 2.03~ ($\tau$)~ 
    &  ~2.39~ ($\Sigma$) & ~ 2.52~ ($\tau$)~ &  ~ 2.61~ ($\tau$)~ & ~ 3.08~ ($\tau$)~  \\
    & $\mu_x $ & ~ 0.00~ & ~ 0.00~ &~ 0.33 ~ & ~ 0.00~ & ~ 3.35~ & ~ 0.94~ & ~ 5.11~ \\
    & $\mu_y $ & ~ 0.71~ & ~ 0.47~ &~ 0.00 ~ & ~ 0.27~ & ~ 0.00~ & ~ 0.00~ & ~ 0.00~ \\
    &          &         &         &         &         &   &         &           \\

3.0 &   gap    & ~ 1.05~ ($\Sigma$) ~ & ~ 1.92~ ($\Sigma$) ~ &~ 2.09~ ($\tau$) ~ 
    & ~ 2.36~ ($\tau$) ~ & ~ 2.40~ ($\Sigma$) ~ &  ~ 2.72~ ($\tau$) ~ & ~ 3.12~ ($\tau$) ~\\
    & $\mu_x $ & ~ 0.00~ & ~ 0.00~ &~ 0.08 ~ & ~ 3.82~ & ~ 0.00~ &  ~ 2.02~ & ~ 3.32~ \\
    & $\mu_y $ & ~ 1.07~ & ~ 0.68~ &~ 0.00 ~ & ~ 0.00~ & ~ 0.29~ &  ~ 0.00~ & ~ 0.00~ \\
    &          &         &         &         &         &     &       &          \\ 

4.0 &   gap    & ~ 1.04~ ($\Sigma$) ~ & ~ 1.89~ ($\Sigma$) ~ &~ 2.11~ ($\tau$) ~ 
    & ~ 2.24~ ($\tau$) ~ & ~ 2.45~ ($\Sigma$) ~ &  ~ 2.80~ ($\tau$) ~ & ~ 3.12~ ($\tau$) ~ \\ 
    & $\mu_x $ & ~ 0.00~ & ~ 0.00~ &~ 1.52 ~ & ~ 3.63~ & ~ 0.00~  &  ~ 2.64~ & ~ 1.21~  \\
    & $\mu_y $ & ~ 1.32~ & ~ 1.17~ &~ 0.00 ~ & ~ 0.00~ &~ 0.21~ &  ~ 0.00~ & ~ 0.00~ \\
\hline
\end{tabular}
\caption {Energy gaps from the lowest triplet state and the 
corresponding transition dipole moments (Debye) in Tetracene and 
substituted tetracene as a function of $\epsilon$. The Even and odd spaces 
under the $C_2 \times e$-$h$ symmetry are labelled  $\Sigma$ and $\tau$, 
respectively. All energies are in eV.}
\label{tab-trip}
\end{table}
\end{center}

    We have compared the singlet-triplet gaps, $E_{S0}-E_{T1}$,
$E_{S0}-E_{T2}$ and singlet-singlet gap, $E_{S0}-E_{S1}$ in \ref{tab-s1t1t2},
as a function of site energy, $\epsilon$. The singlet-triplet gap for the 
unsubstituted tetracene is 1.22 eV which compares well with the experimental 
value of 1.25 eV \cite{stgap-tetra}.
The triplet or spin gap slightly decreases from 1.22 eV for unsubstituted 
tetracene to 1.17 eV for $\epsilon = 2.0$ eV, 1.10 and 1.02 eV for 
$\epsilon = 3.0$ and $4.0$ eV, respectively. In all these cases, the triplet 
gap is less than half of the lowest singlet gap. 

\begin{center}
\begin{table} 
\setlength{\tabcolsep}{2.0pt}
\begin{tabular}{|c|l|l|l|} \hline
$\epsilon$ & ~~~ $T_1$  & ~~~ $T_2$ & ~~~ $S_1$  \\ \hline
0.0 &   ~ 1.22~ &  ~ 2.27~ & ~ 3.18 ~   \\
2.0 &  ~ 1.17 ~ & ~ 2.23 ~ & ~ 2.72~   \\
3.0 &  ~ 1.10 ~ & ~ 2.15 ~ & ~ 2.45~   \\
4.0 &  ~ 1.02 ~ & ~ 2.06 ~ & ~ 2.20~   \\ \hline
\end{tabular}
\caption{Energy levels of  $T_1$, $T_2$ and $S_1$, for the tetracene molecule
as a function of site energy, $\epsilon$. All energies are in eV.}
\label{tab-s1t1t2}
\end{table}
\end{center}

We note from \ref{tab-s1t1t2} that the energy of the $S_1$ state is higher 
than twice the energy of the $T_1$ state, in all cases. Thus the first 
condition 
for singlet fission (SF) is satisfied by both unsubstituted and substituted 
tetracenes. In the case of unsubstituted tetracenes, the two-photon state is 
below the one-photon state and two-photon energy is 3.04 eV and this is also
more than twice the triplet gap of 1.22 eV. As the donor-acceptor strength is 
increased, $S_1$ energy reduces and for $\epsilon = 4.0 $ eV, the $S_1$ 
energy is 2.20 eV against a $T_1$ energy of 1.02 eV. The smaller difference 
in energy between $S_1$ and twice $T_1$  energy implies that less energy 
is lost to heat  in the SF process. Thus strong donor-acceptor substituted 
tetracenes have an edge over weakly substituted tetracenes. The $T_2$ state 
energy of weakly substituted tetracenes  is well below the $S_1$ state. But, 
for strongly substituted tetracenes, $E(T_2)$  is only 0.14 eV below the 
$S_1$ state. These calculations are in the gas phase and intermolecular 
interactions are expected to reduce the $S_1$ energy more than 
$T_2$ energy and could therefore lead to $E(T_2) > E(S_1)$. The $T_1$ 
energies are slightly more than 1 eV implying that the open cell voltage of 
OPVC will be in the desired range.

\subsection{Charge and Spin Density}

\begin{figure}
\begin{center}
\vspace{0.5cm}
\includegraphics[height=6.0cm,width=12cm]{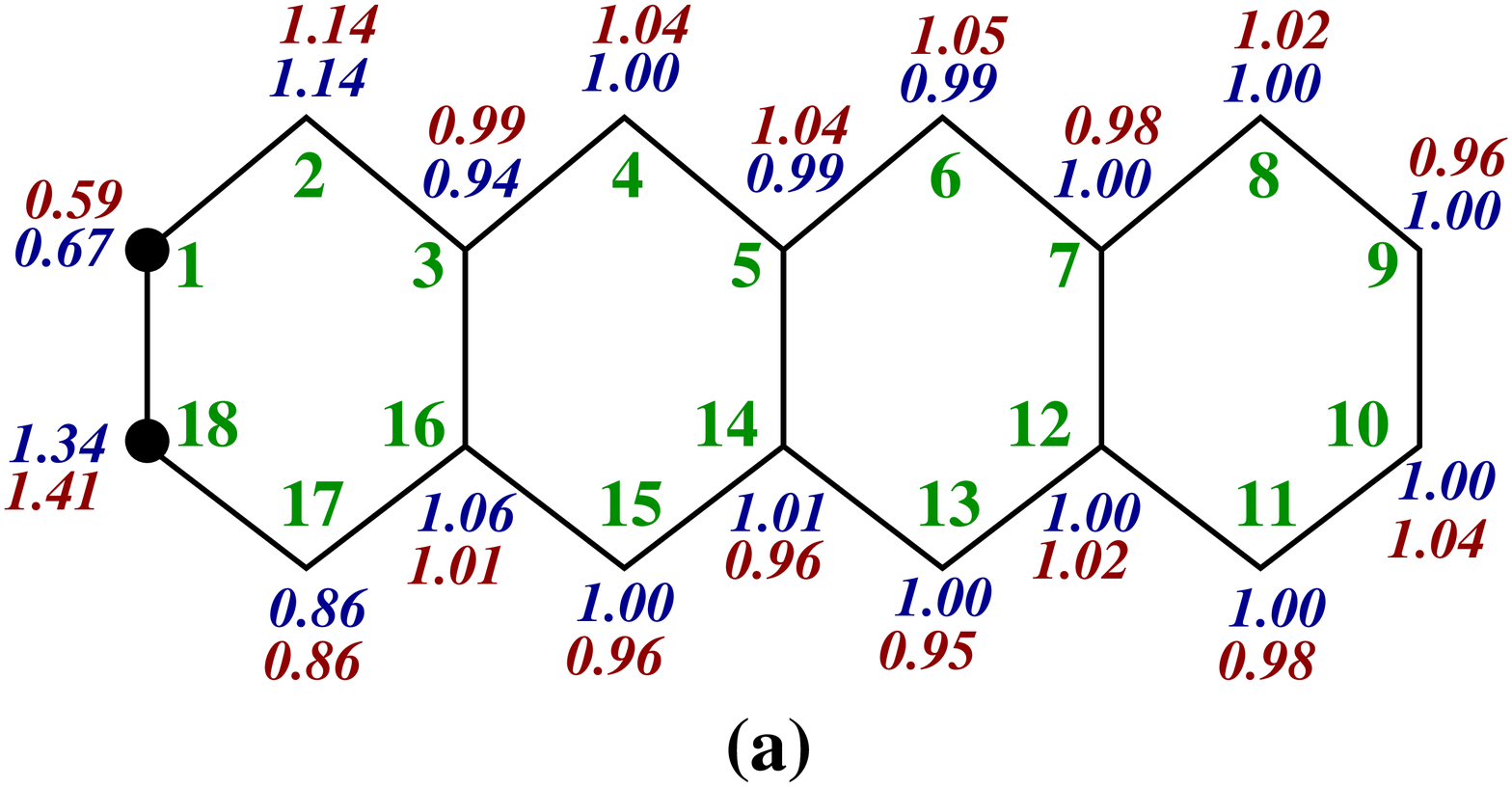} \\
\vspace{0.5cm}
\includegraphics[height=6.0cm,width=12cm]{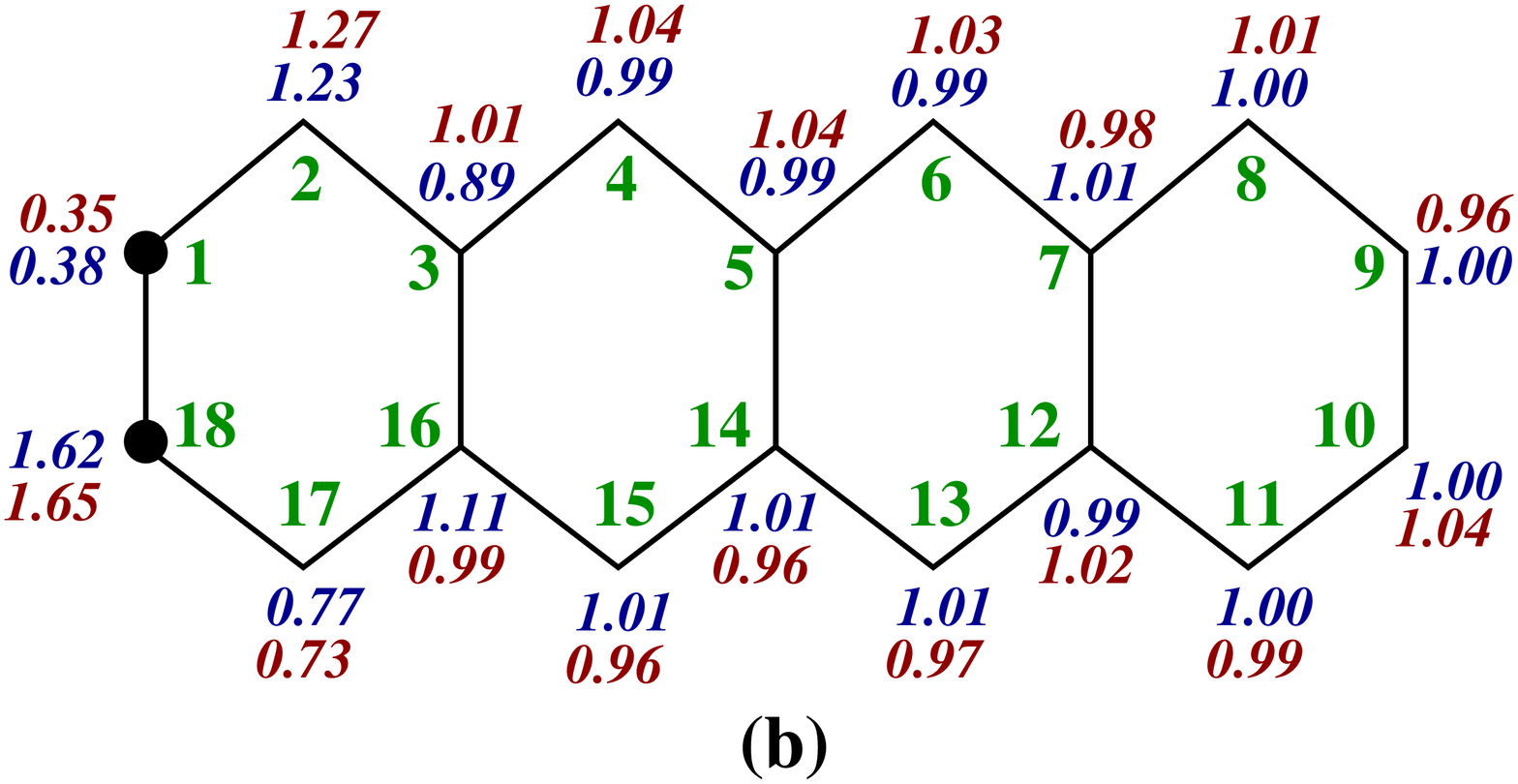} \\
\vspace{0.5cm}
\end{center}
\caption{Charge density for  ground state (numbers in blue) and optically 
allowed state (numbers in red, $T_2$ state), as a function of site 
energy, $\epsilon$. (a) for $\epsilon = 2.0$ eV and (b) for 
$\epsilon = 4.0$ eV. Numbers inside the ring in green represent the 
site/orbital indices.}
\label{chr-trip-fig}
\end{figure}

The charge densities in triplet states of unsubstituted tetracene are uniform.
In the substituted tetracenes, we have shown the charge densities for $T_1$ 
and the triplet state to which transition is most intense ( $T_6$ for 
$\epsilon = 2.0$ eV and  $T_5$ for $\epsilon = 3.0 eV, 4.0$ eV), 
in \ref{chr-trip-fig} (The charge density data for $\epsilon = 3.0$ eV is given
in supporting information). The charge densities in the  $T_1$ state for 
all cases show large variation from the mean near the site of substitution. 
However, unlike in the case of singlets, the charge density fluctuation is 
more extended, although the 
change is largest near the site of substitution. In the triplet state with 
largest transition dipole to the  $T_1$ state, the difference in charge 
density compared to  $T_1$ is much smaller than in the case of singlets. 

\begin{figure}
\begin{center}
\vspace{0.5cm}
\includegraphics[height=6.0cm,width=12cm]{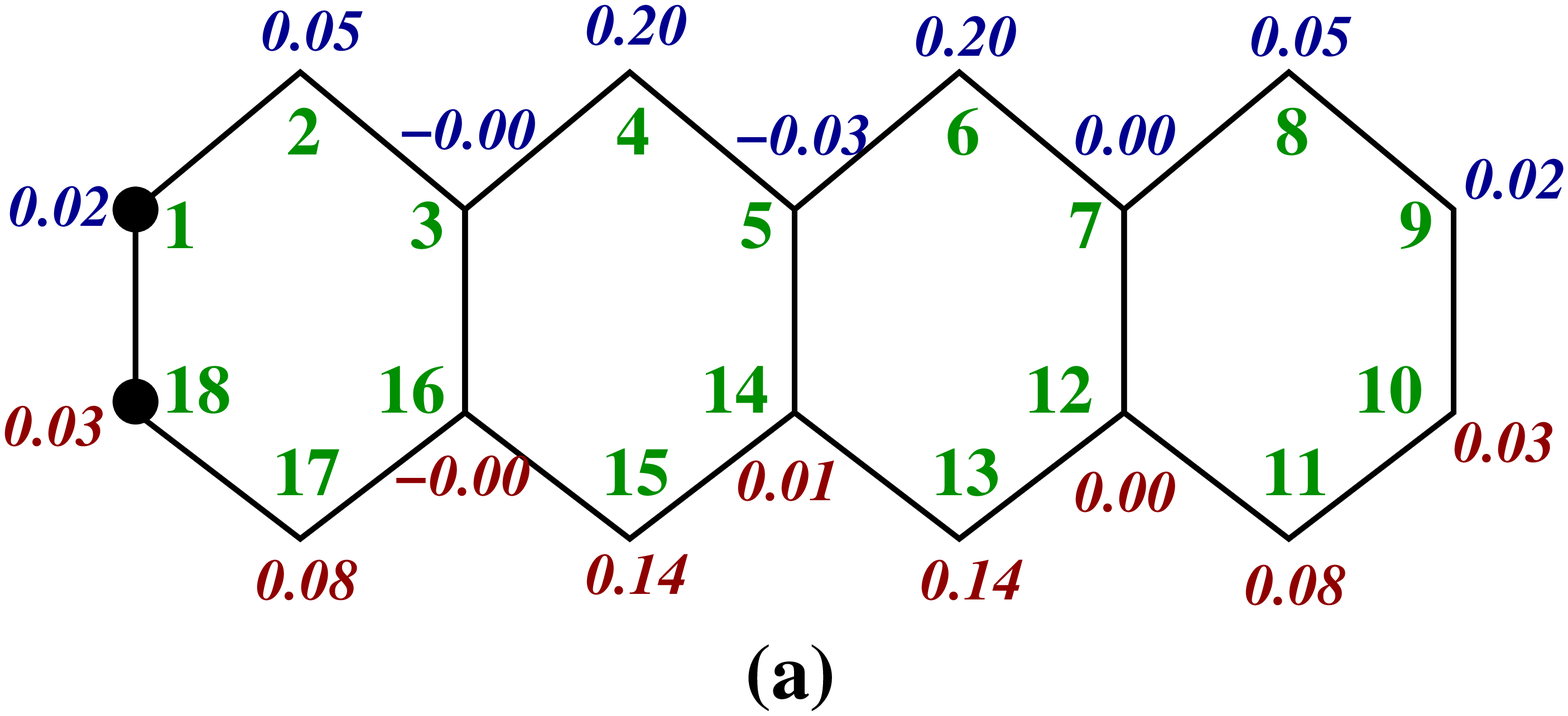} \\
\vspace{0.5cm}
\includegraphics[height=6.0cm,width=12cm]{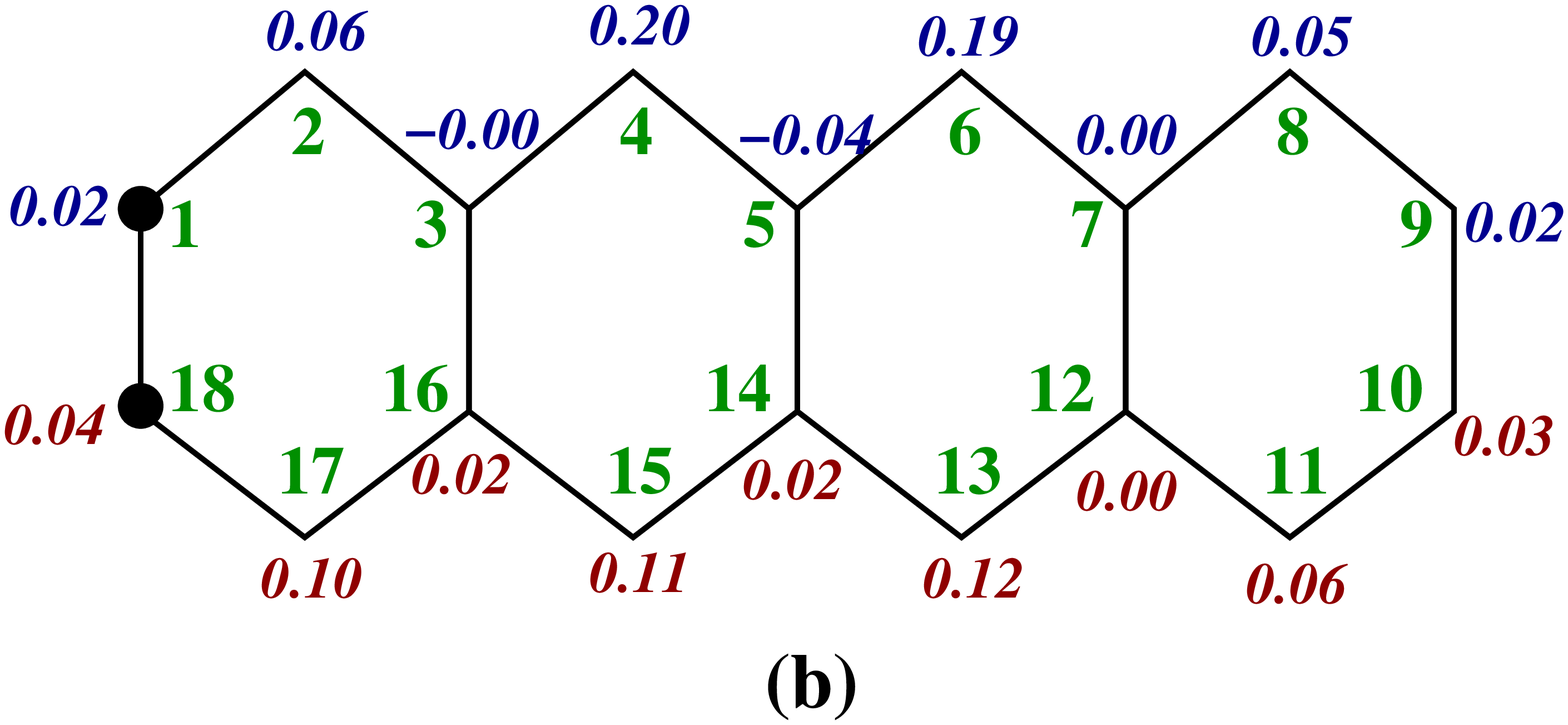} \\
\vspace{0.5cm}
\includegraphics[height=6.0cm,width=12cm]{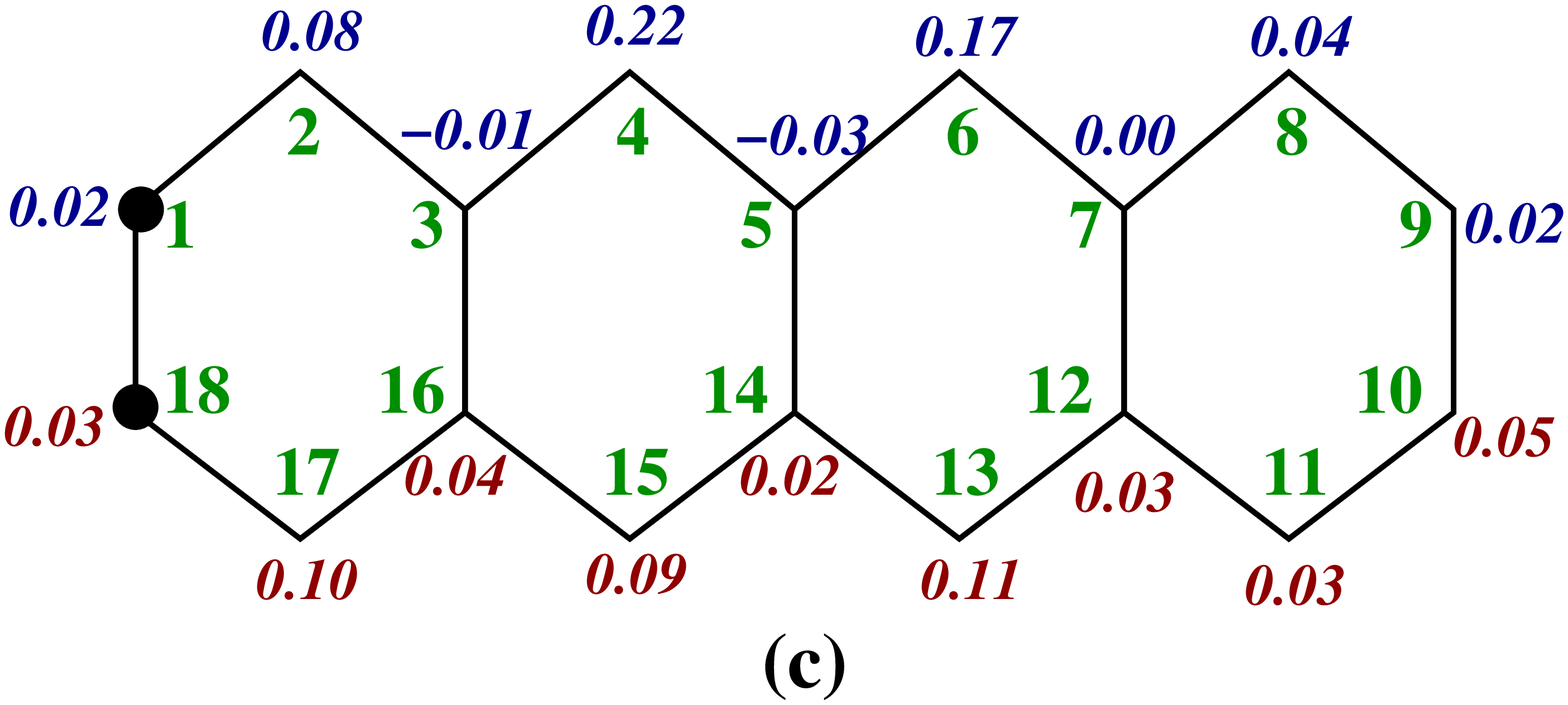} \\
\vspace{0.5cm}
\end{center}
\caption{Spin density for  $T_1$ state (numbers in blue) and optically 
allowed state ($T_2$, numbers in red), as a function of site energy, 
$\epsilon$. (a) for $\epsilon = 0.0$ eV, (b) for $\epsilon = 2.0$ eV and 
(c) for $\epsilon = 4.0$ eV. Numbers inside the ring in green represent 
the site/orbital indices. $C_2$ symmetry about the long axis is valid for
spin densities. }
 
\label{spn-trip-fig}
\end{figure}

We have also computed the spin densities in the  $T_1$ state and the most 
strongly  dipole allowed excited state in both substituted and unsubstituted 
tetracenes (see \ref{spn-trip-fig}). Eventhough in substituted  tetracenes, 
the $C_2$ symmetry about the long axis is broken, the spin densities retain 
this symmetry. This is 
because the donor and acceptors have the same substitution strength and spin 
densities of holes and electrons are the same. The spin densities are all 
positive except, mainly at sites 5 and 14, eventhough the magnitudes are 
rather small. The positive spin densities are larger at the interior of 
tetracene (carbon sites 4, 6, 13 and 15). The spin density magnitudes are 
rather weakly dependent on the strength of substitution. In the excited  
triplet states, there are no sites with negative spin densities and the spin 
densities are more uniform, reflecting higher kinetic energy in the state 
due to greater spin blocking of the delocalization.

\section{Nonlinear properties}

\begin{figure}
\begin{center}
\includegraphics[height=10.0cm,width=8cm]{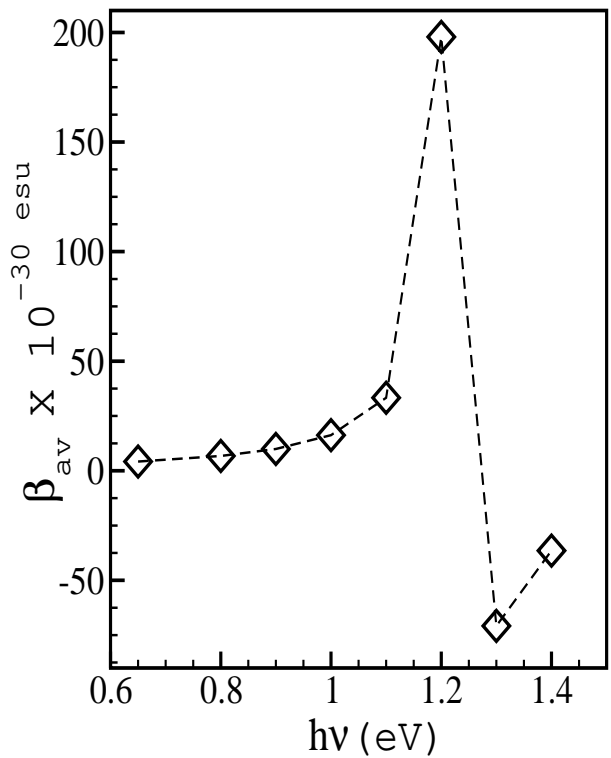} \\
\caption{Dependence of the norm of $||\vec{\beta}_{av}||$ on frequency for
tetracene for site energy, $\epsilon = 3.0$ eV. }
\label{nlo-fig}
\end{center}
\end{figure}

  We have computed the linear polarizability $\alpha_{ij} (-\omega,\omega)$ for 
tetracene and substituted tetracene and 
the second harmonic generation (SHG) coefficients, 
$\beta_{ijk} (-2\omega; \omega,\omega)$ for substituted tetracenes, at a
frequency corresponding to  0.65 eV. We have employed the correction vector 
method, which includes all excitations of the model Hamiltonian; the method
has been described in detail earlier \cite{sr-nlo}. We have tabulated only 
the non-zero and unique components of polarizabilities in \ref{tab-nlo}. We 
note from  \ref{tab-nlo} that, $\alpha_{xx}$ remains almost independent 
of substitution strength as the substituents are placed symmetrically 
about the molecular axis
(see \ref{mol-struct-fig}). The $\alpha_{yy}$ component increases with the  
substitution strength. The SHG coefficient $\beta_{xxx}$ is zero, while 
$\beta_{xxy}$ and $\beta_{xyx}$ are equal by permutation symmetry. However, 
$\beta_{xxy}$ is not equal to $\beta_{yxx}$, because of the substitution along 
the Y-axis. The $\beta$ components are in general small, and increases only 
for strong substitution. The $\beta_{yyy}$ component is negative for weak 
substitution but changes sign and becomes comparable to $\beta_{xxy}$ for 
strong substitution strength. Although we are not close to resonance 
at $\omega = 0.65$ eV excitation frequency, the negative sign 
of $\beta_{yyy}$ implies that some states with large transition dipoles 
between excited states have a sign
opposite to that of transition dipole with the ground state. These studies
indicate that the substituted tetracenes are not good as SHG molecules.
The $||\vec{\beta}_{av}||$ value
is nearly doubled as the strength of D/A is increased. The excitation energy
decreases as we increase the D/A strength while the transition dipole moment
increases (see \ref{tab-sing}), which leads to higher $||\vec{\beta}_{av}||$ with 
the increase in D/A strength. We have also given a plot of 
$\vec{\beta}_{av}$ as a function of the laser excitation frequency in 
\ref{nlo-fig} for site energy, $\epsilon = 3.0$ eV. We find that only
near the resonance $||\vec{\beta}_{av}||$ has a high value of about
 $192 \times 10^{-30}$ esu 
and the resonance occurs at $E_g/2$, as expected.

\begin{center}
\begin{table} 
\setlength{\tabcolsep}{2.0pt}
\begin{tabular}{|c|c|l|l|l|l|l|l|} \hline
~ $ \epsilon$ (in eV) ~ & ~~ $\alpha_{xx}$ ~ & ~~ $\alpha_{yy}$ ~ & ~~ $\alpha_{av}$ ~ &  
~~ $ \beta_{xxy}$ ~ & ~~~ $\beta_{yxx}$ ~ & ~~~ $\beta_{yyy}$ ~ & ~~~ $\beta_{av}$ ~  \\  \hline
~ 0.0 ~ &  ~ 3.77 ~ & ~ 1.77 ~ & ~ 1.85~ &  
~ 0.00 ~ & ~~ 0.00 ~~ & ~~~ 0.00 ~ & ~~ 0.00 ~ \\
~ 1.0 ~ & ~ 3.63 ~ & ~ 1.71 ~ & ~ 1.79 ~ &
~ 0.86 ~ & ~~ 1.73 ~ & ~~ -0.43 ~ & ~~ 0.43 ~ \\ 

~ 2.0 ~ & ~ 3.66 ~ & ~ 1.80 ~ & ~ 1.81 ~ &
~ 2.16 ~ & ~~ 3.89 ~ & ~~ -1.30 ~ & ~~ 1.73 ~ \\ 

~ 3.0 ~ & ~ 3.67 ~ & ~ 1.94 ~ & ~ 1.87 ~ &
~ 3.89 ~ & ~~ 7.34 ~ & ~~ -0.86 ~ & ~~ 4.32 ~ \\ 

~ 4.0 ~ & ~ 3.66 ~ & ~ 2.12 ~ & ~ 1.93 ~ &
~ 6.04 ~ & ~11.66 ~ & ~~~ 1.30 ~ & ~~ 9.50 ~ \\ 

~ 5.0 ~ & ~ 3.63 ~ & ~ 2.28 ~ & ~ 1.97 ~ &
~ 8.20 ~ & ~16.84 ~ & ~~~ 7.77 ~ & ~18.99 ~ \\ 

~ 6.0 ~ & ~ 3.59 ~ & ~ 2.42 ~ & ~ 2.00 ~ &
~ 9.93 ~ & ~21.58 ~ & ~ 18.99 ~ & ~32.81 ~ \\ 

\hline
\end{tabular}
\caption{First order polarizability $\alpha_{ij}(-\omega;\omega)$ and 
first order hyperpolarizability, $\beta_{ijk}(-2\omega;\omega,\omega)$, 
as a function of site energy, $\epsilon$, at $\omega=0.65$ eV. All quantities are in
e.s.u.}
\label{tab-nlo}
\end{table}
\end{center}

\section{Summary}
Tetracene and substituted tetracenes are important functional molecules. 
Obtaining reliable low-lying electronic excited states is a major challenge.
We have employed  the VB method to obtain the singlet and triplet states of
the molecules within  PPP model. The triplet space dimension is more than
901 million while the dimensionality of the space spanned by the singlets is
nearly 450 million. Our studies show that the strongly substituted tetracenes
can be useful in organic photovoltaics as they satisfy the energy criteria for 
singlet fission. The changes in equilibrium geometries of the excited states
relative to the ground states are small implying that the Stark shifts will
be small. Thus, the excitation energies   are close to their value in 
equilibrium geometries. The spin density in triplets are mainly confined to
the middle of the ring while charge densities of triplets and singlets are 
large at the substituted sites. The exact SHG coefficients computed for 
substituted tetracenes show that the SHG response of these molecules 
are small. 

\begin{acknowledgement} We thank Department of Science and Technology, India
 for financial support (DST0691). Y.A.P. thanks Disha Programme for women 
in Science, (DST01226) for financial support. 
\end{acknowledgement}

\noindent
\begin{suppinfo}

We have given the following figures for site energy $\epsilon = 3.0$ eV. \\
\noindent
i) Charge densities for ground state and dipole allowed vertically excited 
state.
 
\noindent
ii) Difference in charge densities from ground state to vertically
excited states in even space and odd space.

\noindent
iii) Bond order for ground state and dipole allowed vertically excited 
state.

\noindent
iv) Difference in bond orders from ground state to vertically
excited states in even space  and odd space.

\noindent
v) Charge densities for lowest triplet state and dipole
allowed vertically excited state. 

\noindent
vi) Spin densities for lowest triplet state  and dipole allowed vertically 
excited state.
 
\end{suppinfo}

\end{document}